\documentclass[12pt]{article}
\usepackage[english]{babel}
\usepackage{amsmath}
\usepackage{amssymb}
\usepackage{bbm}
\usepackage{color}
\usepackage{multicol}
\usepackage{graphicx}
\usepackage{epsfig}
\usepackage[titletoc,toc,title]{appendix}
\Roman{section}
\newcommand{\eq}[1]{\begin{equation}#1\end{equation}}
\newcommand{\naw}[1]{\left(#1\right)}
\newcommand{\ket}[1]{\left|#1\right>}
\newcommand{\bra}[1]{\left<#1\right|}
\newcommand{\av}[1]{\left<#1\right>}

\newcommand{\modu}[1]{\left|#1\right|}
\newcommand{\poisson}[1]{\left\{#1\right\}}

\begin{document}

\begin{center}
\textsc{\Large{Examining the effect of quantum strategies on symmetric
conflicting interest games}}
\newline

\large{Katarzyna Bolonek-Laso\'n}\footnote{kbolonek@uni.lodz.pl}\\ 
\emph{\normalsize{Faculty of Economics and Sociology, Department of Statistical Methods, \\University of Lodz,
41/43 Rewolucji 1905 St., 90-214 Lodz,  Poland.}}\\
\end{center}

\begin{abstract}
The explicit construction is presented of two-player game satisfying: (\textit{i}) symmetry with respect to the permutation of the players; (\textit{ii}) the existence of upper bound on total payoff following from Bell inequality; (\textit{iii}) the existence of unfair equilibrium with total payoff saturating the above bound. The quantum counterpart of the game is considered which possesses only fair equilibria and strategies outperforming the classical ones.
\end{abstract}

\section{Introduction}
Probably the most striking feature of quantum theory is nonlocality, i.e. the existence of correlations which are not admissible in local theory. The correlations which are admitted by a local theory must satisfy a set of inequalities (Bell inequalities) which can be violated by quantum mechanical correlations \cite{Bell}. The violation of Bell inequalities has been confirmed experimentally \cite{Aspect}.

Nonlocality, expressed in terms of violation of Bell inequalities, is inherent to quantum theory and appears to be useful in practice (see \cite{Buhrman} and the references contained therein).

Bell inequalities can be also discussed in the context of game theory. In particular, in order to relate the nonlocality to the advantages of the strategies based on quantum correlations, a quantum version of the game with incomplete information \cite{Harsanyi} has been proposed \cite{Cheon1} and further developed in a number of papers \cite{Iqbal}, \cite{Flitney1}, \cite{Iqbal1}, \cite{Hill}. Recently, Brunner and Linden \cite{Brunner} considered more general setting for quantum Bayesian games where nonlocal resources provide an advantage over any classical strategy. The examples of games presented in Ref. \cite{Brunner} are the games of common interest. On the other hand, Pappa et al. \cite{Pappa} considered a particular example of two-player conflicting interest game where quantum mechanics also offers an advantage over the classical strategies. The game they considered is a combination of the Battle of Sexes and CHSH games. Further examples of conflicting interest games where quantum mechanics offers an advantage have been given by Situ \cite{Situ}. Roy et al. \cite{Roy}, by  a slight modification of the utility functions proposed in \cite{Pappa}, constructed the examples of games where quantum strategies can outperform the unfair classical equilibrium ones. 

The notion of quantum conflicting interest games can be extended to the multiplayer case. The existence of various forms of entanglement in multipartite systems makes the problem of the relation between nonlocality and the advantage coming from quantum strategies more complicated. For example, Situ et al. \cite{Situ1} constructed an example of Bayesian three-player game based on Svetlichny inequality \cite{Svetlichny}. In this case the advantages of the quantum game are based on the correlations that can be reduced to the mixtures of two-player quantum ones related locally to the third player.

An example of conflicting interest three-player game based on Bell inequalities has been given by the author \cite{Bolonek}. In that paper the scheme for constructing such games was outlined. However, the general formulae for three-player case are quite complicated. On the other hand, the two-player case can be described more explicitly. It is the aim of the present paper to provide such description which generalizes the particular example considered by Pappa et al. \cite{Pappa}.

We shall follow the notion of conflicting interest game used by Pappa et al. \cite{Pappa} (cf. also \cite{osborne}) which is defined in terms of Nash equilibria. Consider a game played by two players, $A$ and $B$, possessing at least two Nash equilibria such that the corresponding payoffs obey $F_A^{(I)}>F_B^{(I)}$ and  $F_A^{(II)}<F_B^{(II)}$. Moreover, let us assume that for any fair equilibrium, $F_A^{(f)}=F_B^{(f)}$ (if it exists), $F_A^{(f)}<F_A^{(I)}$ and $F_B^{(f)}<F_B^{(II)}$. Then each player prefers a different equilibrium. In this sense (cf. Refs. \cite{Pappa} and \cite{osborne}) we are speaking about a conflicting interest game.

 We are interested in the advantages over classical strategies offered by quantum mechanics. In particular, we study in more general setting the question raised in Ref. \cite{Pappa}: can the genuinely quantum strategies convert the conflicting interest game into the common interest one.

A particularly interesting situation arises when the initial game is symmetric with respect to the permutation of the players. Then the occurrence of unfair equilibria breaks this symmetry. This situation resembles the phenomenon of spontaneous symmetry breaking which plays an important role in statistical physics and quantum field theory. There the dynamics (i.e. the Hamiltonian) is invariant under the action of some symmetry group while the ground/equilibrium state breaks this symmetry. It can also happen that a symmetry broken at the classical level is restored when quantum effects are taken into account. The simplest example is the onedimensional double well which is parity invariant but the classical equilibrium (ground) states break parity. The quantum mechanical tunelling between both wells restores symmetry. We shall see that something similar happens here. This analogy is, obviously, quite far but still interesting and nice.

The paper is organized as follows. Section II is devoted to the construction of general game with incomplete information and conflicting interest. Its quantum counterpart is considered in Sec.\,III. Finally, some conclusions are described in the last section. The more technical details are relegated to the Appendices.

\section{Two-player Bayesian games with conflicting interest}

We consider the following situation. There are two players, Alice (A) and Bob (B) and each player can acquire a type $x_i$, $i\in\poisson{A,B}$, $x_i\in\poisson{0,1}$, according to the probability distribution $P(\underline{x})\equiv P(x_A,x_B)$. They decide on their actions $y_i$, $y_i\in\poisson{0,1}$, according to a chosen strategy. The expected payoff of  each player reads
\begin{equation}
F_i=\sum_{(\underline{x},\underline{y})}P(\underline{x})p(\underline{y}|\underline{x})u_i(\underline{x},\underline{y})\label{a}
\end{equation}
with $p\naw{\underline{y}|\underline{x}}\equiv p\naw{y_A,y_B|x_A,x_B}$ being the probability the players choose actions $\underline{y}\equiv\naw{y_A,y_B}$ given their types were $\underline{x}\equiv\naw{x_A,x_B}$; $u_i\naw{\underline{x},\underline{y}}$ are the utility functions determining the gains of players depending on their types and actions.

Let us discuss the general constraints imposed on the formula (\ref{a}). The notions of fair and unfair equilibria are set in a proper framework if we assume that our game is symmetric with respect to the permutation of the players. As a result we have the following symmetry relations:
\begin{equation}
u_A\naw{x_A,x_B,y_A,y_B}=u_B\naw{x_B,x_A,y_B,y_A}.\label{a1}
\end{equation}
What concerns the probabilities we assume that they obey the no-signalling conditions 
\begin{equation}
\begin{split}
& \sum_{y_B}p\naw{y_A,y_B|x_A,x_B}=\sum_{y_B}p\naw{y_A,y_B|x_A,x_B'}\\
& \sum_{y_A}p\naw{y_A,y_B|x_A,x_B}=\sum_{y_A}p\naw{y_A,y_B|x_A',x_B}
\end{split}\label{r4}
\end{equation}
together with the normalization ones
\eq{\sum_{\underline{y}}p\naw{\underline{y}|\underline{x}}=1\;\text{ for all } \; \underline{x}.\label{r5}}
With the above conditions we consider two types of probability distributions:
\begin{itemize}
\item[(i)] \underline{the classical ones}: one assumes further constraints in the form of Bell inequalities \cite{Clauser}.
In order to write out these constraints in terms of probabilities $p\naw{\underline{y}|\underline{x}}$ entering the payoffs (\ref{a}) we ascribe two pairs of observables, $A_x$, $B_x$, $x\in\poisson{0,1}$, to the players $A$ and $B$, respectively. They are indexed by the players types and can acquire two values, $A_x=\pm 1$, $B_x=\pm 1$. In terms of expectation values of these observables the Bell inequality reads \cite{Clauser}
\begin{equation}
\modu{\av{A_0B_0}+\av{A_1B_0}+\av{A_0B_1}-\av{A_1B_1}}\leq 2.\label{r1}
\end{equation} 
By defining the relevant probabilities as 
\begin{equation}
p\naw{y_A,y_B|x_A,x_B}=p\naw{A_{x_A}=2y_A-1\wedge B_{x_B}=2y_B-1}
\end{equation}
one can rewrite (\ref{r1}) in the form
\begin{equation}
\begin{split}
&\Big|\sum_{y_A,y_B=0}^1\naw{2y_A-1}\naw{2y_B-1}\big (p\naw{y_A,y_B|0,1}+p\naw{y_A,y_B|1,0}+\\
&\qquad +p\naw{y_A,y_B|0,0}-p\naw{y_A,y_B|1,1}\big )\Big |\leq 2.
\end{split}\label{r2}
\end{equation}
Assuming the inequality (\ref{r1}) (and, equivalently, (\ref{r2})) together with the ones obtained from (\ref{r1}) by all possible replacements $0\leftrightarrow 1$ of the players types one finds, by virtue of Fine's theorem \cite{Fine}$\div$\cite{Halliwell1}, the following representation of relevant probabilities in terms of hidden variables $\lambda$:
\eq{p\naw{y_A,y_B|x_A,x_B}=\int \text{d}\lambda\, \rho\naw{\lambda}p_A\naw{y_A|x_A,\lambda}p_B\naw{y_B|x_B,\lambda}\label{s1},}
$\rho\naw{\lambda}$ being the probability distribution of $\lambda$. Since there are only two possible actions per player it is sufficient to consider only hidden variables providing two bits so that $p_A\naw{y_A|x_A,\lambda}=p_A\naw{y_A|x_A,\lambda_A}$, $p_B\naw{y_B|x_B,\lambda}=p_B\naw{y_B|x_B,\lambda_B}$.
In game-theoretic language the players receive advice from a classical source (advisor) characterized by $\rho\naw{\lambda}$. The probability $p\naw{\underline{y}|\underline{x}}$ factorizes provided the players are insensitive to the advisor suggestions, $p_A\naw{y_A|x_A,\lambda}=p_A\naw{y_A|x_A}$ etc.
It is worth to notice that Fine's theorem states also that Bell inequalities imply the existence of joint probability distribution for all four observables $A_x$, $B_x$ which yields the probabilities $p\naw{\underline{y}|\underline{x}}$ as marginals. Therefore, the game is classical in the sense that probabilities enter here in the same way as in all classical systems where it is allowed to consider joint probability distributions for any set of observables. From this point of view the only additional constraint imposed is that of no-signalling.
\item[(ii)] \underline{the quantum case}: a quantum source/advisor is characterized by a choice of twopartite density matrix $\rho$. 
In order to define the relevant probabilities one chooses again two pairs of observables, $A_x$ and $B_x$, $x\in\poisson{0,1}$, which are hermitean operators acting in twodimensional Hilbert spaces of the players and admit the spectral decompositions
\eq{\begin{split}
& A_x=1\cdot A_x^1+(-1)\cdot A_x^0,\quad \mathbbm{1}=A_x^1+A_x^0\\
& B_x=1\cdot B_x^1+(-1)\cdot B_x^0,\quad \mathbbm{1}=B_x^1+B_x^0
\end{split}}
with $A_x^y$, $B_x^y$ being the corresponding projections. As a result we get the following expressions for the payoffs:
\eq{F_i=\sum_{\underline{x},\underline{y}}P\naw{\underline{x}}\text{Tr}\naw{\rho\naw{A_{x_A}^{y_A}\otimes B_{x_B}^{y_B}}}u_i\naw{\underline{x},\underline{y}}.\label{a4}}
The general form of $A_x$ and $B_x$ reads 
\eq{A_x=\vec{n}_x^A\cdot\vec{\sigma},\quad B_x=\vec{n}_x^B\cdot\vec{\sigma}\label{a6}}
with $\vec{n}_x^A$, $\vec{n}_x^B$ being the unit vectors, $\vec{n}_x^A=\naw{\sin\theta_x^A\cos\varphi_x^A,\sin\theta_x^A\sin\varphi_x^A,\cos\theta_x^A}$ and the similar formula for $\vec{n}_x^B$.
Note that in the quantum case the Bell inequalities are, in general, violated.
\item[(iii)] \underline{the superquantum case}: there exist non-signalling distributions which are not of quantum mechanical origin \cite{Popescu}. We will not consider such probability distributions. 
\end{itemize}

We shall also assume that the distribution of the player types is uniform,
\eq{P(\underline{x})=\frac{1}{4}.\label{r3}}
In order to construct our game we start with the utility functions $u_i\naw{\underline{x},\underline{y}}$ defined in Table \ref{t1}.

\begin{table}[h!]
\caption{The utilities of players}
 \begin{small}
\begin{tabular}{|c|c|c|c|c|c|}\cline{3-6}
\multicolumn{2}{c}{}& \multicolumn{2}{|c|}{$x_B=0$} & \multicolumn{2}{|c|}{$x_B=1$}\\
\cline{3-6}
\multicolumn{2}{c|}{} & $y_B=0$ & $y_B=1$ & $y_B=0$ & $y_B=1$  \\
\hline
$x_A=0$ & $y_A=0$ & $\naw{s_1,t_1}$ & $\naw{s_2,t_2}$ & $\naw{s_5,t_5}$ & $\naw{s_6,t_6}$  \\
\cline{2-6}
 & $y_A=1$ & $\naw{s_3,t_3}$ & $\naw{s_4,t_4}$ & $\naw{s_7,t_7}$ & $\naw{s_8,t_8}$  \\
  \hline
$x_A=1$ & $y_A=0$ &  $\naw{s_9,t_9}$ & $\naw{s_{10},t_{10}}$ & $\naw{s_{13},t_{13}}$ & $\naw{s_{14},t_{14}}$ \\
\cline{2-6}
 & $y_A=1$ & $\naw{s_{11},t_{11}}$ & $\naw{s_{12},t_{12}}$ & $\naw{s_{15},t_{15}}$ & $\naw{s_{16},t_{16}}$  \\
  \hline
 \end{tabular}\end{small}\label{t1}
\end{table}

 We impose the following constraints:
\begin{itemize}
\item[a)] $u_i\naw{\underline{x},\underline{y}}$ obey the symmetry conditions following from eqs. (\ref{a1}) 
\eq{\begin{split}
\begin{array}{llll}
s_7=t_{10}, & s_{10}=t_7, & s_2=t_3, & s_3=t_2,\\
s_5=t_9, & s_9=t_5, & s_{11}=t_6, & s_6=t_{11},\\
s_{12}=t_8, & s_8=t_{12}, & s_{15}=t_{14}, & s_{14}=t_{15},\\
s_{1}=t_1, & s_4=t_{4}, & s_{13}=t_{13}, & s_{16}=t_{16},
\end{array}
\end{split}\label{c}}

\item[b)]  the total payoff $F_A+F_B$ is expressible solely in terms of the combination of probabilities entering the Bell inequality (\ref{r1}). 
We impose this condition for the following reasons. As it has been already mentioned the Bell inequality (\ref{r1}) together with the remaining three obtained by exchanging the observable indices form necessary and sufficient conditions for the probabilities to be genuinely classical, i.e. resulting as marginals from the joint probability distribution for all observables. It cannot be violated on classical level. Therefore, the total payoff is bounded from above by virtue of eq. (\ref{r1}) by some value $F$. If there exist Nash equilibria saturating the bound they can be either fair, with $F_A=F_B=\frac{1}{2}F$, or unfair, $F_A\neq  F_B=F-F_A$; in the latter case there must exist, due to the symmetry of the game, the accompanying equilibrium with $\widetilde{F}_A=F_B=F-F_A$, $\widetilde{F}_B=F_A$. Assuming, for example, $2F_A>F$ we find that Alice favorizes the first unfair equilibrium while Bob - the second one. So if there exists an unfair equilibrium saturating the bound for total payoff we are dealing with conflicting interest game.

\qquad In order to find the resulting constraint on $s_i$ and $t_i$, $i=1,\ldots,16$, we express the sum $F_A+F_B$ in terms of probabilities $p(\underline{y},\underline{x})$; to this end we use eqs. (\ref{a}) and (\ref{r3}) together with the data entering Table \ref{t1}. Then we demand that $p(\underline{y}|\underline{x})$ enter $F_A+F_B$ in the combination appearing on the left hand side of (\ref{r2}) and, as a result, we find the following constraints (cf. the Appendix A for details): 
\eq{
\begin{split}
& s_1-s_2+t_1-t_2=s_5-s_6+t_5-t_6\\
& s_1-s_2+t_1-t_2=s_9-s_{10}+t_9-t_{10}\\
& s_1-s_2+t_1-t_2=s_{14}-s_{13}+t_{14}-t_{13}\\
& s_4-s_1+t_4-t_1=0\\
& s_3-s_2+t_3-t_2=0\\
& s_8-s_5+t_8-t_5=0\\
& s_7-s_6+t_7-t_6=0\\
& s_{12}-s_9+t_{12}-t_9=0\\
& s_{11}-s_{10}+t_{11}-t_{10}=0\\
& s_{16}-s_{13}+t_{16}-t_{13}=0\\
& s_{14}-s_{15}+t_{14}-t_{15}=0.\\
\end{split}\label{a5}}

Note that we could generalize the condition (b) by demanding that $F_A+F_B$ is expressible in terms of some linear combination of probabilities entering all four Bell inequalities.
\end{itemize}
 The general solution of (a) and (b) depends  on the number of free parameters and is given in Table \ref{t2}.

\begin{table}[h!]
\caption{The utilities of players after using eqs. (\ref{c}) and (\ref{a5})}
 \begin{small}
\begin{tabular}{|c|c|c|c|}\cline{3-4}
\multicolumn{2}{c}{}& \multicolumn{2}{|c|}{$x_B=0$}\\
\cline{3-4}
\multicolumn{2}{c|}{} & $y_B=0$ & $y_B=1$ \\
\hline
$x_A=0$ & $y_A=0$ & $\naw{s_1,s_1}$ & $\naw{s_2,s_3}$ \\
\cline{2-4}
 & $y_A=1$ & $\naw{s_3,s_2}$ & $\naw{s_1,s_1}$ \\
  \hline
$x_A=1$ & $y_A=0$ &  $\naw{s_9,s_5}$ & $\naw{-2s_1+s_2+s_3+s_5-s_7+s_9,s_7}$\\
\cline{2-4}
 & $y_A=1$ & $\naw{-2s_1+s_2+s_3+s_5-s_6+s_9,s_6}$ & $\naw{s_{5}-s_8+s_9,s_{8}}$ \\
  \hline
 \multicolumn{2}{|c}{} &\multicolumn{2}{|c|}{$x_B=1$}\\
\hline
$x_A=0$ & $y_A=0$ & $\naw{s_5,s_9}$ & $\naw{s_6,-2s_1+s_2+s_3+s_5-s_6+s_9}$ \\
\cline{2-4}
 & $y_A=1$ & $\naw{s_7,-2s_1+s_2+s_3+s_5-s_7+s_9}$ & $\naw{s_8,s_5-s_8+s_9}$ \\
  \hline
$x_A=1$ & $y_A=0$ &  $\naw{s_{13},s_{13}}$ & $\naw{s_{14}, 2s_1-s_2-s_3+2s_{13}-s_{14}}$\\
\cline{2-4}
 & $y_A=1$ & $ \naw{2s_1-s_2-s_3+2s_{13}-s_{14},s_{14}}$ & $\naw{s_{13},s_{13}}$ \\
  \hline  
\end{tabular}\end{small}\label{t2}
\end{table}

Now, we must select some candidate for unfair Nash equilibrium, i.e. the corresponding set of probabilities $p\naw{\underline{y}|\underline{x}}$. We consider "pure" strategies, i.e. the ones obeying $p(\underline{y}|\underline{x})=0,1$. For definiteness we take 
\eq{p\naw{\underline{y}|\underline{x}}=\delta_{y_A,0}\,\delta_{y_B,1-x_B};\label{a3}}
where $\delta_{y,x}$ is the Kronecker delta function. \\
Let us stress that eq. (\ref{a3}) is a specific classical strategy that we impose as a candidate for Nash equilibrium which determines the utility table (Table 2). Had we chosen a different one it would have led to a different utility table. 
Obviously, the probability distribution defined by eq. (\ref{a3}) obeys no-signalling and normalization conditions, eqs. (\ref{r4}) and (\ref{r5}), respectively. It saturates the Bell inequality (\ref{r2}) so the bound on total payoff $F_A+F_B$ is also saturated. It is one of $2^3$ extremal points of the convex set defined by eqs. (\ref{r4}) and (\ref{r5}). All  strategies corresponding to these extremal points are called the pure ones. Let us now impose the next constraint:
\begin{itemize}
\item[c)] eq. (\ref{a3}) defines unfair Nash equilibrium such that the total payoff $F_A+F_B$ saturates the upper bound following from Bell inequality. In order to derive the relevant conditions on the utility functions it is sufficient to consider only the subset of pure strategies. In fact, any probability distribution describing pure strategy and defining Nash equilibrium in this subset continues to describe a Nash equilibrium if also mixed strategies are admitted \cite{osborne}. This is because our payoffs are linear functions of probabilities $p(\underline{y},\underline{x})$ so they acquire their extrema on extremal points of the convex set defined by eqs. (\ref{r2}) and (\ref{r3}). The existence of additional constraint in form of Bell inequalities only strenghtens this argument. Obviously, extending the set of strategies by including the mixed ones can produce new Nash equilibria. However, if a new equilibrium is a fair one, the payoff of at least one player must be smaller than that corresponding to the strategy (\ref{a3}) because the total payoff of the latter saturates the bound following from Bell inequality.
\end{itemize}

Demanding that eq. (\ref{a3}) defines unfair Nash equilibrium in the set of pure strategies with total payoff saturating classical bound yields the following conditions: 
\eq{\begin{split}
& s_2+s_5+s_{10}+s_{13}>s_2+s_5+s_{12}+s_{15},\\
& s_2+s_5+s_{10}+s_{13}>s_4+s_7+s_{10}+s_{13},\\
& s_2+s_5+s_{10}+s_{13}>s_4+s_7+s_{12}+s_{15},\\
& t_2+t_5+t_{10}+t_{13}>t_1+t_5+t_9+t_{13},\\
& t_2+t_5+t_{10}+t_{13}>t_1+t_6+t_9+t_{14},\\
& t_2+t_5+t_{10}+t_{13}>t_2+t_6+t_{10}+t_{14},\\
& s_2+s_5+s_{10}+s_{13}>t_2+t_5+t_{10}+t_{13},\\
& s_1-s_2+t_1-t_2<0.
\end{split}\label{bac}}

 They express the property that both Alice and Bob have nothing to gain by changing unilateraly her/his strategy (while $F_A+F_B$ saturates the relevant bound).
Taking into account the relations (\ref{c}) and (\ref{a5})  we find finally
\eq{\begin{split}
& 2s_2+2s_3+s_8+s_{14}>4s_1+s_7+s_{13},\\
& s_2+s_5>s_1+s_7,\\
& 3s_2+2s_3+s_5+s_8+s_{14}>5s_1+2s_7+s_{13},\\
& s_3+s_7>s_1+s_5,\\
& s_3+s_6+s_7+s_{14}>s_1+2s_5+s_{13},\\
& s_6+s_{14}>s_5+s_{13},\\
& s_2+s_5>s_1+s_7,\\
& 2s_1<s_2+s_3.
\end{split}\label{a7}}

Let us notice that the above conditions imply that, as it has been already explained, we are dealing with conflicting interest game. 

To conclude, we have defined a wide family of two-players symmetric games with incomplete information and conflicting interest. The main assumption we made is that the total payoff for unfair Nash equilibrium saturates the upper bound following from Bell inequality; this leads automatically to the game with conflicting interest.

\section{The quantum counterpart of Bayesian game with conflicting interest}
Once the utilities $u_i\naw{\underline{x},\underline{y}}$ are known we can construct the quantum counterpart of our game. To this end we use eq. (\ref{a4}). Let us first note an important new feature which emerges on the quantum level if we are going to take advantage  from the possibility of violating the Bell inequalities. In the previous section 
we considered the classical game in full generality, i.e. the probabilities $p\naw{\underline{y}|\underline{x}}$ were assumed to obey no conditions except the no-signalling ones and the Bell inequalities; according to Fine's theorem the latter are, however, equivalent to the, natural on the classical level, condition that there exists a joint probability distribution for all random variables under consideration \cite{Fine}, \cite{Fine1}. Obviously, a further choice of probability distribution $\rho(\lambda)$ and the functions $p\naw{y|x,\lambda}$ would impose, according to eq. (\ref{s1}), additional constraints on the players strategies. However, all consideration relied on Bell inequalities only so there is no necessity for making further assumptions.

On the quantum level we would like to construct an advisor acting according to the quantum rules which allow to outperform classical strategies. In other words, we should violate Bell inequalities which constraint the effectiveness of classical strategies. Not all quantum states lead to their violation. Therefore, we have to select a particular state $\rho$ of the advisor. This choice restricts the set of allowed quantum strategies. The properties of the game depend not only on the utility functions but also on the players strategies admitted. We shall see below that, with the particular $\rho$ selected, one can make the payoff functions of both players equal. The resulting game has then only fair equilibria. The twofold role of quantum strategies which both raise the payoffs in fair equilibria and eliminate the unfair ones seems to be not sufficiently stressed in literature. 

In order to construct the appropriate game we take   
\eq{\rho=\ket{\Psi}\bra{\Psi}\label{ba1}}
where 
\eq{\ket{\Psi}=\frac{1}{\sqrt{2}}\naw{\ket{00}+\ket{11}}.\label{b3}}
Note that
\eq{e^{\frac{i\chi_1}{2}\sigma_3}\otimes e^{\frac{i\chi_2}{2}\sigma_3}\ket{\Psi}=(-1)^n\ket{\Psi}\label{b2}}
provided $\chi_1+\chi_2=2\pi n$.  
The quantum strategies are characterized, according to eqs. (\ref{a6}), by eight angles, $\naw{\varphi_1,\theta_1}\equiv\naw{\varphi_0^A,\theta_0^A}$, $\naw{\varphi_2,\theta_2}\equiv\naw{\varphi_1^A,\theta_1^A}$, $\naw{\varphi_3,\theta_3}\equiv\naw{\varphi_0^B,\theta_0^B}$ and $\naw{\varphi_4,\theta_4}\equiv\naw{\varphi_1^B,\theta_1^B}$.
By inserting eqs. (\ref{a6}), (\ref{r3}) and (\ref{b3}) into eq. (\ref{a4}) one obtains explicit expressions for quantum payoffs. They are quite involved and will be not written out explicitly. However, it can be checked that the Alice and Bob payoffs coincide, $F_A=F_B\equiv F$, provided one additional simple constraint is imposed 
\eq{2s_1-s_2-s_3-s_5+s_6+s_7-s_8=0.\label{pl}}
We consider this constraint as a part of the definition of our game. Let us note that our aim is to show that quantum strategies can outperform the classical ones and not that they have to. The additional condition (\ref{pl}) is added for this purpose.

The equality $F_A=F_B$ implies that in the quantum case we are dealing with fair equilibria only. Moreover, they correspond to the maxima of the common payoff function $F$ or, equivalently, to the maxima of total payoff $F_A+F_B\equiv 2F$. Now, $F$ is a linear function of the combination of correlation functions entering the left-hand side of Bell inequality (\ref{r1}). According to Tsirelson \cite{Tsirelson} the latter is bounded from above by $2\sqrt{2}$. It is well known that this bound can be saturated by considering the observables built out of $\sigma_1$ and $\sigma_2$ only. Therefore, we can look for the maxima of $F$ (i.e. the Nash equilibria of quantum game) by imposing additional constraints $\theta_i=\frac{\pi}{2}$, $i=1,\ldots,4$. Then, using eqs. 
(\ref{c}) and (\ref{a5}) we find (see Appendix B for details): 
\eq{
\begin{split}
& F\equiv F_A=F_B=\frac{1}{16}\left[(2s_1-s_2-s_3)\left(\cos\naw{\varphi_1+\varphi_3}+\cos\naw{\varphi_2+\varphi_3}+\right.\right.\\
&\left.\left.+\cos\naw{\varphi_1+\varphi_4}-\cos\naw{\varphi_2+\varphi_4}\right)+\naw{2s_2+2s_3+4s_5+4s_9+4s_{13}}\right]\end{split}\label{b1}} 
Due to the last inequality (\ref{a7}), $2s_1-s_2-s_3<0$, we have to find the minimal value of the combination of cosine functions entering the right hand side of eq. (\ref{b1}). Now, according to eq. (\ref{b2}), the payoff functions (\ref{b1}) are invariant under the transformations $\varphi_{1,2}\rightarrow\varphi_{1,2}+\chi_1$, $\varphi_{3,4}\rightarrow\varphi_{3,4}+\chi_2$, $\chi_1+\chi_2=2n\pi$. Therefore, there exist the whole families of maxima of $F$.  They read
\eq{
\begin{split}
& \varphi_1+\varphi_2=\frac{\pi}{4}+\frac{\naw{3n+r+s+2}\pi}{2}\\
& \varphi_2+\varphi_3=-\frac{\pi}{4}+\frac{\naw{n-r+3s+2}\pi}{2}\\
& \varphi_1+\varphi_4=-\frac{\pi}{4}+\frac{\naw{n+3r-s+2}\pi}{2}
\end{split}}
provided 
\eq{n+3r-s=4k\quad\text{or}\quad 4k+1;}
the payoffs of both players read 
\eq{F_A=F_B=\frac{1}{16}\naw{2\sqrt{2}\naw{s_2+s_3-2s_1}+2\naw{s_2+s_3}+4\naw{s_5+s_9+s_{13}}}\label{ab}}
while the total payoff is given by
\begin{equation}
F_A+F_B=2F_A=\frac{1}{8}\naw{2\sqrt{2}\naw{s_2+s_3-2s_1}}+\frac{1}{4}\naw{s_2+s_3+2s_5+2s_9+2s_{13}}\label{as}
\end{equation}
We conclude that the quantum counterpart of our game possesses only fair equilibria corresponding to the payoffs (\ref{ab}) provided the additional constraint (\ref{pl}) has been imposed.
The total payoff (\ref{as}) saturates the Tsirelson bound, i.e. it is a global maximum of the total payoff function. 

Let us remind that the maximal total payoff on classical level, resulting fom Bell inequality, reads
\eq{F_A+F_B=\frac{1}{4}\naw{s_2+s_3-2s_1}+\frac{1}{4}\naw{s_2+s_3+2s_5+2s_9+2s_{13}}.}
Quantum strategies outperform the classical ones.

\section{Conclusion}
We have constructed a general classical two-player game with incomplete information and conflicting interest. To this end we imposed the following conditions: (\textit{i}) the game is symmetric with respect to the permutation of the players; (\textit{ii}) the total payoff is expressible in terms of Bell operator; (\textit{iii}) there exists an unfair Nash equilibrium saturating the bound on total payoff resulting from Bell inequality.

Then we proposed a quantum counterpart of the game. To this end we picked out a particular model of quantum advisor based on the pure state (\ref{b3}). It appeared that, by adding one additional condition on utilities (eq. (\ref{pl})), we were able to make payoff functions equal. This implies immediately that our quantum game possesses only fair equilibria. Moreover, the quantum strategies outperform, due to the violation of Bell inequality, the classical ones. The quantum equilibria saturate the Tsirelson bound; the quantum character of the game is maximally exploited. 

Let us stress again that the probabilities entering the classical payoff functions are restricted only by no-signalling condition. The quantum ones, on the other hand, are constrained by a specific form of quantum advisor which is chosen in such a way that the Bell inequality can be violated. This constraint on  probabilities excludes the unfair equilibria.  The quantum game is no longer a conflicting interest one. This property seems to be not stressed sufficiently in the literature. 

Note that the general solution to our conditions (\ref{a5}), (\ref{bac}) and  (\ref{pl}) spans a ninedimensional variety. The example considered by Pappa et al. \cite{Pappa} corresponds to the following particular choice of the parameters:
\begin{displaymath}
\begin{split}
& s_1=0, s_2=1, s_3=\frac{1}{2}, s_4=0, s_5=0, s_6=1, s_7=\frac{1}{2}, s_8=0,\\
&  s_9=0, s_{10}=1, s_{11}=\frac{1}{2}, s_{12}=0, s_{13}=\frac{3}{4}, s_{14}=0, s_{15}=0, s_{16}=\frac{3}{4}\\
& t_1=0, t_2=\frac{1}{2}, t_3=1, t_4=0, t_5=0, t_6=\frac{1}{2}, t_7=1, t_8=0,\\
&  t_9=0, t_{10}=\frac{1}{2}, t_{11}=1, t_{12}=0, t_{13}=\frac{3}{4}, t_{14}=0, t_{15}=0, t_{16}=\frac{3}{4}.
\end{split}
\end{displaymath}

\begin{appendices}
\section{Derivation of eqs. (\ref{a5})}
In order to derive the conditions (\ref{a5}) we start with eqs. (\ref{a}). Using the notation described in Table 1 we write out explicitly the expected payoffs of both players. For example, one obtains (note that $P\naw{x}=\frac{1}{4}$, cf. eq. (\ref{r3}))
\begin{equation}
\begin{split}
& F_A=\frac{1}{4}\left( s_1p\naw{00|00}+s_2p(01|00)+s_3p(10|00)+s_4p(11|00)+\right.\\
& \qquad +s_5p(00|01)+s_6p(01|01)+s_7p(10|01)+s_8p(11|01)+\\
& \qquad +s_9p(00|10)+s_{10}p(01|10)+s_{11}p(10|10)+s_{12}p(11|10)+\\
& \qquad \left. +s_{13}p(00|11)+s_{14}p(01|11)+s_{15}p(10|11)+s_{16}p(11|11)\right)
\end{split}\label{ap5}
\end{equation}
and similar equation for $F_B$ with $s_i$ replaced by $t_i$, $i=1,\ldots,16$.\\
Now, the relevant Bell inequality reads
\eq{\modu{\av{A_0B_0}+\av{A_1B_0}+\av{A_0B_1}-\av{A_1B_1}}\leq 2\label{ap7}}
where $A_k$ $(B_k)$ refers to Alice (Bob) observables, $k=0,1$ denotes the type and $A_k$ $(B_k)$ acquire the values 1 if $y_A=1$ $(y_b=1)$ or -1 if $y_A=0$ $(y_B=0)$. From the very definition of correlation function one finds (i,j=0,1):
\eq{\av{A_iB_j}=p(00|ij)+p(11|ij)-p(01|ij)-p(10|ij)\label{ap1}}

Eqs. (\ref{ap1}), together with the normalization conditions
\eq{p(00|ij)+p(11|ij)+p(01|ij)+p(10|ij)=1}
allow us to express some probabilities $p(\underline{y}|\underline{x})$ in terms of relevant correlation functions and the remaining probabilities which can be considered as independent ones
\eq{p(00|ij)=\frac{1}{2}\naw{1-2p(11|ij)+\av{A_iB_j}}\label{ap3}}
\eq{p(01|ij)=\frac{1}{2}\naw{1-2p(10|ij)-\av{A_iB_j}}\label{ap4}}
Inserting (\ref{ap3}) and (\ref{ap4}) into (\ref{ap5}) one finds
\eq{
\begin{split}
& F_A=\frac{1}{4}\left(\naw{\frac{s_1-s_2}{2}}\av{A_0B_0}+\naw{\frac{s_5-s_6}{2}}\av{A_0B_1}+\naw{\frac{s_9-s_{10}}{2}}\av{A_1B_0}+\right.\\
&\qquad + \naw{\frac{s_{13}-s_{14}}{2}}\av{A_1B_1}+\naw{s_3-s_2}p(10|00)+(s_4-s_1)p(11|00)+\\
& \qquad +(s_7-s_6)p(10|01)+(s_8-s_5)p(11|01)+(s_{11}-s_{10})p(10|10)+\\
& \qquad +(s_{12}-s_9)p(11|10)+(s_{15}-s_{14})p(10|11)+(s_{16}-s_{13})p(11|11)+\\
&\qquad + \naw{\frac{s_1+s_2+s_5+s_6+s_9+s_{10}+s_{13}+s_{14}}{2}};
\end{split}\label{ap6}}
repeating the same procedure for $F_B$ and adding the resulting expression to (\ref{ap6}) we obtain the formula for the total payoff $F_A+F_B$. It takes the form of linear combination of terms proportional (with coefficients depending on $s_i$'s and $t_i$'s) either to correlation functions $\av{A_iB_k}$, $i,k=0,1$, or to the independent probabilities $p(10|x_Ax_B)$ and $p(11|x_Ax_B)$ plus a free term depending on $s_i$'s
 and $t_i$'s only. By demanding that the independent probabilities do not enter the final result while the correlation functions enter only in the combination appearing on the left hand side of eq. (\ref{ap7}) one finds the relations (\ref{a5}).
 
 As it has been noticed in the main text one could generalize the above reasoning by demanding that $F_A+F_B$ is expressible in terms of linear combination of the expressions appearing on the left hand side of (\ref{ap7}) and the remaining Bell inequalities
 \eq{\modu{\av{A_1B_0}+\av{A_0B_0}+\av{A_1B_1}-\av{A_0B_1}}\leq 2}
 \eq{\modu{\av{A_0B_1}+\av{A_1B_1}+\av{A_0B_0}-\av{A_1B_0}}\leq 2}
 \eq{\modu{\av{A_1B_1}+\av{A_0B_1}+\av{A_1B_0}-\av{A_0B_0}}\leq 2.}
 
\section{Derivation of eq. (\ref{b1})}

 The observables $A_x$ and $B_x$ entering the quantum payoffs are given by eqs. (\ref{a6}). Therefore, we get for the type $"x"$:
 \begin{equation}
 A_x=\left(\begin{array}{cc}
 \cos\theta_x^A & \sin\theta_x^A e^{-i\varphi_x^A}\\
 \sin\theta_x^A e^{i\varphi_x^A} & -\cos\theta_x^A
 \end{array}\right ),\qquad x=0,1\label{ap8}
 \end{equation}
  \begin{equation}
 B_x=\left(\begin{array}{cc}
 \cos\theta_x^B & \sin\theta_x^B e^{-i\varphi_x^B}\\
 \sin\theta_x^B e^{i\varphi_x^B} & -\cos\theta_x^B
 \end{array}\right ).\label{ap9}
 \end{equation}
 As it has been explained in the main text  we can restrict ourselves to the case $\theta_i^{A,B}=\frac{\pi}{2}$. Even within this restricted set of observables the upper quantum bound for Bell correlations (Tsirelson bound \cite{Tsirelson}) can be achieved. According to our assumption (\ref{ap8}) and (\ref{ap9}) reduce to 
 \begin{equation}
 A_x=\left(\begin{array}{cc}
 0 & e^{-i\varphi_x^A}\\
 e^{i\varphi_x^A} & 0
 \end{array}\right ),\qquad 
 B_x=\left(\begin{array}{cc}
 0 &  e^{-i\varphi_x^B}\\
  e^{i\varphi_x^B} & 0
 \end{array}\right ).\label{ap10}
 \end{equation}
 The spectral projectors for the observables (\ref{ap10}) read
 \begin{equation}
 \begin{split}
 & A_x^0=\frac{1}{2}\left(\begin{array}{cc}
 1 & -e^{-i\varphi_x^A}\\
 -e^{i\varphi_x^A} & 1 
 \end{array}\right ),\qquad   A_x^1=\frac{1}{2}\left(\begin{array}{cc}
 1 & e^{-i\varphi_x^A}\\
 e^{i\varphi_x^A} & 1 
 \end{array}\right )\\
 & B_x^0=\frac{1}{2}\left(\begin{array}{cc}
 1 & -e^{-i\varphi_x^B}\\
 -e^{i\varphi_x^B} & 1 
 \end{array}\right ),\qquad   B_x^1=\frac{1}{2}\left(\begin{array}{cc}
 1 & e^{-i\varphi_x^B}\\
 e^{i\varphi_x^B} & 1 
 \end{array}\right )
 \end{split}\label{ap11}
 \end{equation}
 From eqs. (\ref{ap11}) and (\ref{ba1}), (\ref{b3}) we easily find:
 \eq{\text{Tr}\naw{\rho\naw{A_{x_A}^{y_A}\otimes B_{x_B}^{y_B}}}=\bra{\Psi}A_{x_A}^{y_A}\otimes B_{x_B}^{y_B}\ket{\Psi}}
 or 
  \eq{
  \begin{split}
  & \text{Tr}\naw{\rho\naw{A_{x_A}^{y_A}\otimes B_{x_B}^{y_B}}}=\frac{1}{2}\left(\bra{0}A_{x_A}^{y_A}\ket{0}\bra{0} B_{x_B}^{y_B}\ket{0}+\bra{0}A_{x_A}^{y_A}\ket{1}\bra{0} B_{x_B}^{y_B}\ket{1}+\right.\\
  & \hspace{4cm} +\left. \bra{1}A_{x_A}^{y_A}\ket{0}\bra{1} B_{x_B}^{y_B}\ket{0}+\bra{1}A_{x_A}^{y_A}\ket{1}\bra{1} B_{x_B}^{y_B}\ket{1}\right)
  \end{split}\label{ap12}}
Eqs. (\ref{ap11}) and (\ref{ap12}) imply
\eq{
\begin{split}
& \text{Tr}\naw{\rho\naw{A_{x_A}^{0}\otimes B_{x_B}^{0}}}=\frac{1}{4}\naw{1+\cos\naw{\varphi_{x_A}^A+\varphi_{x_B}^B}}\\
& \text{Tr}\naw{\rho\naw{A_{x_A}^{0}\otimes B_{x_B}^{1}}}=\frac{1}{4}\naw{1-\cos\naw{\varphi_{x_A}^A+\varphi_{x_B}^B}}\\
& \text{Tr}\naw{\rho\naw{A_{x_A}^{1}\otimes B_{x_B}^{0}}}=\frac{1}{4}\naw{1-\cos\naw{\varphi_{x_A}^A+\varphi_{x_B}^B}}\\
& \text{Tr}\naw{\rho\naw{A_{x_A}^{1}\otimes B_{x_B}^{1}}}=\frac{1}{4}\naw{1+\cos\naw{\varphi_{x_A}^A+\varphi_{x_B}^B}}.
\end{split}}
According to the convention adopted below eq.\,(\ref{b2}) we put $\varphi_0^A\equiv\varphi_1$, $\varphi_1^A\equiv\varphi_2$, $\varphi_0^B\equiv\varphi_3$, $\varphi_1^B\equiv\varphi_4$. Eq. (\ref{a4}) and the Table \ref{t1} yield now the following expressions for the payoffs
\eq{
\begin{split}
& F_A=\frac{1}{16}\left(\cos(\phi_1+\phi_3)(s_1-s_2-s_3+s_4)+\cos(\phi_1+\phi_4)(s_5-s_6-s_7+s_8)+\right.\\
& \qquad +\cos(\phi_2+\phi_3)(s_9-s_{10}-s_{11}+s_{12})+\cos(\phi_2+\phi_4)(s_{13}-s_{14}-s_{15}+s_{16})+\\
& \qquad + s_1+s_2+s_3+s_4+s_5+s_6+s_7+s_8+s_9+s_{10}+s_{11}+s_{12}+s_{13}+s_{14}+\\
& \qquad \left. +s_{15}+s_{16}\right)
\end{split} \label{ap13}}
\eq{
\begin{split}
& F_B=\frac{1}{16}\left(\cos(\phi_1+\phi_3)(t_1-t_2-t_3+t_4)+\cos(\phi_1+\phi_4)(t_5-t_6-t_7+t_8)+\right.\\
& \qquad +\cos(\phi_2+\phi_3)(t_9-t_{10}-t_{11}+t_{12})+\cos(\phi_2+\phi_4)(t_{13}-t_{14}-t_{15}+t_{16})+\\
& \qquad + t_1+t_2+t_3+t_4+t_5+t_6+t_7+t_8+t_9+t_{10}+t_{11}+t_{12}+t_{13}+t_{14}+\\
& \qquad \left. +t_{15}+t_{16}\right)
\end{split} \label{ap14}}
Taking into account the relations described in Table \ref{t2} one can reduce eqs. (\ref{ap13}) and (\ref{ap14}) to
\eq{
\begin{split}
&F_A=\frac{1}{16}\left[ (\cos(\phi_1+\phi_3)+2\cos(\phi_2+\phi_3)-\cos(\phi_2+\phi_4))(2s_1-s_2-s_3)+\right.\\
& \qquad + (\cos(\phi_1+\phi_4)-\cos(\phi_2+\phi_3))(s_5-s_6-s_7+s_8)+\\
& \qquad \left.+2(s_2+s_3+2s_5+2s_9+2s_{13})\right]
\end{split}} 
\eq{
\begin{split}
&F_B=\frac{1}{16}\left[ (\cos(\phi_1+\phi_3)+2\cos(\phi_1+\phi_4)-\cos(\phi_2+\phi_4))(2s_1-s_2-s_3)+\right.\\
& \qquad + (\cos(\phi_2+\phi_3)-\cos(\phi_1+\phi_4))(s_5-s_6-s_7+s_8)+\\
& \qquad \left.+2(s_2+s_3+2s_5+2s_9+2s_{13})\right]
\end{split}}

 \end{appendices}

\subsection*{Acknowledgement}
I am grateful to Prof. Piotr Kosi\'nski  for fruitful discussion and useful remarks.

\end{document}